\pdfoutput=1
\documentclass[useAMS,usenatbib]{mn2e}
\usepackage{apjfonts}
\usepackage{graphicx}
\usepackage{amsmath}
\usepackage{amssymb}
\usepackage{ctable}
\usepackage{fixltx2e}
\usepackage{threeparttable}
\usepackage{hyperref}
\hypersetup{colorlinks=true,linkcolor=blue,citecolor=blue,filecolor=blue,urlcolor=blue}

\newcommand{\be}{\begin{equation}}
\newcommand{\ee}{\end{equation}}
\newcommand{\ba}{\begin{eqnarray}}
\newcommand{\ea}{\end{eqnarray}}

\newcommand{\Rvir}{R_{\rm vir}}

\newcommand{\Mhalo}{{\rm M}_{\rm halo}}

\newcommand{\Ms}{{\rm M}_{\ast}}

\newcommand{\muv}{{\rm M_{UV}}}
\newcommand{\cm}{{\rm cm}}

\newcommand{\pc}{{\rm pc}}

\newcommand{\Myr}{{\rm Myr}}

\newcommand{\Msun}{{\rm M}_{\sun}}
\newcommand{\Zsun}{Z_{\sun}}

\newcommand{\mb}{m_b}

\newcommand{\eb}{\epsilon_b}
\newcommand{\es}{\epsilon_{\ast}}

\newcommand{\nc}{n_{\rm th}}

\newcommand{\LCDM}{$\Lambda$CDM}
\newcommand{\aj}{AJ}
\newcommand{\apj}{ApJ}
\newcommand{\apjl}{ApJ}
\newcommand{\apjs}{ApJS}
\newcommand{\mnras}{MNRAS}
\newcommand{\aap}{A\&A}

\newcommand{\rmxaa}{RMxAA}

\newcommand{\referee}[1]{{\color{black}#1}}

\title[High-$z$ galaxy morphologies and sizes in FIRE-2]
{Simulating galaxies in the reionization era with FIRE-2: morphologies and sizes}

\author[X. Ma et al.]{
  \parbox[t]{1.0\textwidth}{
   Xiangcheng Ma,$^1$\thanks{E-mail: xchma@caltech.edu}
   Philip F. Hopkins,$^1$ 
   Michael Boylan-Kolchin,$^2$ \\
   Claude-Andr{\'e} Faucher-Gigu{\`e}re,$^3$ 
   Eliot Quataert,$^4$
   Robert Feldmann,$^5$ \\
   Shea Garrison-Kimmel,$^1$
   Christopher C. Hayward,$^6$
   Du{\v s}an Kere{\v s}$^7$ \\ and
   Andrew Wetzel$^8$
  }
  \vspace{5pt} \\
  $^1$TAPIR, MC 350-17, California Institute of Technology, Pasadena, CA 91125, USA \\
  $^2$The University of Texas at Austin, Department of Astronomy, 2515 Speedway, Stop C1400, Austin, Texas 78712-1205 \\
  $^3$Department of Physics and Astronomy and CIERA, Northwestern University, 2145 Sheridan Road, Evanston, IL 60208, USA \\
  $^4$Department of Astronomy and Theoretical Astrophysics Center, University of California Berkeley, Berkeley, CA 94720 \\
  $^5$Institute for Computational Science, University of Zurich, Zurich CH-8057, Switzerland \\
  $^6$Center for Computational Astrophysics, Flatiron Institute, 162 Fifth Avenue, New York, NY 10010, USA \\
  $^7$Department of Physics, Center for Astrophysics and Space Sciences, University of California at San Diego, 9500 Gilman Drive, La Jolla, CA 92093 \\
  $^8$Department of Physics, University of California, Davis, CA 95616, USA \\
}

\pagerange{\pageref{firstpage}--\pageref{lastpage}}
\date{Draft version \today}

\begin{document}
\maketitle
\label{firstpage}

\begin{abstract}
We study the morphologies and sizes of galaxies at $z\geq5$ using high-resolution cosmological zoom-in simulations from the Feedback In Realistic Environments project. The galaxies show a variety of morphologies, from compact to clumpy to irregular. The simulated galaxies have more extended morphologies and larger sizes when measured using rest-frame optical B-band light than rest-frame UV light; sizes measured from stellar mass surface density are even larger. The UV morphologies are usually dominated by several small, bright young stellar clumps that are not always associated with significant stellar mass. The B-band light traces stellar mass better than the UV, but it can also be biased by the bright clumps. At all redshifts, galaxy size correlates with stellar mass/luminosity with large scatter. The half-light radii range from 0.01 to 0.2\,arcsec (0.05--1\,kpc physical) at fixed magnitude. At $z\geq5$, the size of galaxies at fixed stellar mass/luminosity evolves as $(1+z)^{-m}$, with $m\sim1$--2. For galaxies less massive than $\Ms\sim10^8\,\Msun$, the ratio of the half-mass radius to the halo virial radius is $\sim10\%$ and does not evolve significantly at $z=5$--10; this ratio is typically 1--5\% for more massive galaxies. A galaxy's `observed' size decreases dramatically at shallower surface brightness limits. This effect may account for the extremely small sizes of $z\geq5$ galaxies measured in the Hubble Frontier Fields. We provide predictions for the cumulative light distribution as a function of surface brightness for typical galaxies at $z=6$. 
\end{abstract}


\section{Introduction}
\label{sec:intro}
High-redshift galaxies are thought to be the dominant source of cosmic reionization \citep[e.g.][]{kuhlen.faucher.2012:reion,robertson.2013:reion.hudf12,robertson.2015:reion.planck}. The number density of these galaxies, as described by the ultraviolet (UV) luminosity function, is well constrained for galaxies brighter than $\rm M_{UV}=-17$ at $z\sim6$ \citep[e.g.][]{mclure.2013:uvlf.z7to9.hudf12,schenker.2013:uvlf.z7to8.udf12,bouwens.2015:uvlf.z4to10,finkelstein.2015:uvlf.combined.field}. Recently, the Hubble Frontier Fields (HFF) program \citep{lotz.2017:hubble.frontier.field}, which takes advantage of strong gravitational lensing by foreground galaxy clusters, has made it possible to estimate UV luminosity functions down to $\rm M_{UV}\sim-13$ \citep[e.g.][]{bouwens.2017:lensing.uncertainty,livermore.2017:faint.galaxies}. But one of the dominant outstanding uncertainties is the intrinsic size distribution of faint galaxies, which is necessary in order to determine the completeness of the observed sample due to surface brightness limits \citep{bouwens.2017:small.galaxy.sizes}.

There are only a few galaxies at these redshifts that have robust size measurements. \citet{oesch.2010:hudf.morphology} measured the sizes of galaxies brighter than $\rm M_{UV}=-19$ at $z=4$--8 in the Hubble Ultra-Deep Field (HUDF) and found that the half-light radii of galaxies  evolve according to $(1+z)^{-m}$, with $m\sim1$--1.5 \citep[see also, e.g.][]{bouwens.2004:hudf.size.evolution,ferguson.2004:size.evolution,ono.2013:size.evolution.udf12,kawamata.2015:hff.size.z6to8}. It is also expected from analytic models that galaxy size decreases with increasing redshift \citep{mo.mao.white.1998:galactic.disk,mo.mao.white.1999:lbg.structure}.

High-redshift galaxies tend to be intrinsically small. The half-light radii of bright galaxies ($\rm M_{UV}<-19$) at $z\sim6$--8 range from 0.5--1\,kpc \citep[e.g.][]{oesch.2010:hudf.morphology}. More recently, \citet{kawamata.2015:hff.size.z6to8} and \citet{bouwens.2017:small.galaxy.sizes} measured the half-light radii for a sample of fainter galaxies ($-19<\rm M_{UV}<-12$) from the HFF. They found that the size--luminosity relation has large scatter, with half-light radii spanning more than an order of magnitude (0.1--1\,kpc) at fixed UV magnitude. A fraction of these faint galaxies have extremely small sizes from a few pc to 100\,pc, although these results are very uncertain because they are far below the resolution of the {\it Hubble Space Telescope (HST)}.

Morphological studies of intermediate-redshift galaxies ($z\sim0.5$--3) reveal that their images typically contain a number of star-forming clumps \citep[e.g.][]{guo.2015:clumpy.galaxy.candels}, which only contribute a small fraction of the total stellar mass \citep[e.g.][]{wuyts.2012:candels.clumps}. So far, the sizes of $z\gtrsim6$ galaxies are measured using noise-weighted stacked images over all available bands (dominated by rest-frame UV), so it is likely that the extremely small galaxy sizes in the HFF are biased by such clumps \citep[e.g.][]{vanzella.2017:globular.formation}. With the {\em James Webb Space Telescope} ({\em JWST}, scheduled to launch in 2018), one will be able to probe these faint, high-redshift galaxies with deeper imaging, higher spatial resolution, and at longer wavelengths. This makes it possible to compare galaxy morphology and sizes in different bands and to recover the stellar mass distribution using multi-band images via pixel-by-pixel spectral energy distribution (SED) modeling \citep[e.g.][]{smith.hayward.2015:sed.modeling}.

The goal of this paper is to make predictions of morphology and sizes for $z\geq5$ galaxies, which can be used to plan and interpret future observations. We use a suite of high-resolution cosmological zoom-in simulations from the Feedback In Realistic Environments (FIRE) project\footnote{http://fire.northwestern.edu} \citep{hopkins.2014:fire.galaxy}. The FIRE simulations include explicit treatments of the multi-phase interstellar medium (ISM), star formation, and stellar feedback. The simulations are evolved using the FIRE-2 code \citep{hopkins.2017:fire2.numerics}, which is an update of the original FIRE code with several improvements to the numerics. These simulations predict stellar mass functions and luminosity functions in broad agreement with observations at these redshifts. When evolved to $z=0$, simulations with the same physics have been shown to also reproduce many other observed galaxy properties \citep[][and references therein]{hopkins.2017:fire2.numerics}.

The paper is organized as follows. In Section \ref{sec:methods}, we describe the simulations briefly and the methods we use to measure galaxy sizes. We present the results in Section \ref{sec:results} and discuss their implications to observations in Section \ref{sec:discussion}. We conclude in Section \ref{sec:conclusion}. We adopt a standard flat {\LCDM} cosmology with {\it Planck} 2015 cosmological parameters $H_0=68 {\rm\,km\,s^{-1}\,Mpc^{-1}}$, $\Omega_{\Lambda}=0.69$, $\Omega_{m}=1-\Omega_{\Lambda}=0.31$, $\Omega_b=0.048$, $\sigma_8=0.82$, and $n=0.97$ \citep{planck.2016:cosmo.param}. We use a \citet{kroupa.2002:imf} initial mass function (IMF) from 0.1--$100\,\Msun$, with IMF slopes of $-1.30$ from 0.1--$0.5\,\Msun$ and $-2.35$ from 0.5--$100\,\Msun$. All magnitudes are in the AB system \citep{oke.gunn.1983:ab.system}.

\section{Methods}
\label{sec:methods}

\subsection{The simulations}
\label{sec:sim}
We use a suite of 15 high-resolution cosmological zoom-in simulations at $z\geq5$ from the FIRE project, spanning a halo mass range $\Mhalo=10^8$--$10^{12}\,\Msun$ at $z=5$. These simulations are first presented in \citet{ma.2017:fire.hiz.smf}. The mass resolution for baryonic particles (gas and stars) is $\mb=100$--$7000\,\Msun$ (more massive galaxies having larger particle mass). The minimum Plummer-equivalent force softening lengths for gas and star particles are $\eb=0.14$--0.42\,pc and $\es=0.7$--2.1\,pc (see table 1 in \citealt{ma.2017:fire.hiz.smf} for details). The softening lengths are in comoving units above $z=9$ but switch to physical units thereafter. All of the simulations are run using exactly identical code {\sc gizmo}\footnote{http://www.tapir.caltech.edu/{\textasciitilde}phopkins/Site/GIZMO.html} \citep{hopkins.2015:gizmo.code}, in the mesh-less finite-mass (MFM) mode with the identical FIRE-2 implementation of star formation and stellar feedback.

The baryonic physics included in FIRE-2 simulations are described in \citet{hopkins.2017:fire2.numerics}, but we briefly review them here. Gas follows an ionized-atomic-molecular cooling curve from $10$--$10^{10}$\,K, including metallicity-dependent fine-structure and molecular cooling at low temperatures and high-temperature metal-line cooling for 11 separately tracked species (H, He, C, N, O, Ne, Mg, Si, S, Ca, and Fe). At each timestep, the ionization states and cooling rates H and He are calculated following \citet{katz.1996:sph.cooling}, with the updated recombination rates from \citet{verner.1996:recombination.rates}, and cooling rates from heavier elements are computed from a compilation of {\sc cloudy} runs \citep{ferland.2013:cloudy.release}, applying a uniform but redshift-dependent photo-ionizing background from \citet{cafg.2009:uvb}, and an approximate model for H\,{\sc ii} regions generated by local sources. Gas self-shielding is accounted for with a local Jeans-length approximation. We do not include a primordial chemistry network nor Pop III star formation, but apply a metallicity floor of $Z=10^{-4}\,\Zsun$.

We follow the star formation criteria in \citet{hopkins.2013:sf.criteria} and allow star formation to take place only in dense, molecular, and locally self-gravitating regions with hydrogen number density above a threshold $\nc=1000\,\cm^{-3}$. Stars form at 100\% efficiency per local free-fall time when the gas meets these criteria, and there is no star formation elsewhere. The galactic-scale star formation efficiency is regulated by feedback to $\sim1\%$ \citep[e.g.][]{orr.2017:fire.ks.law}. The simulations include the following stellar feedback mechanisms: (1) local and long-range momentum flux from radiation pressure, (2) SNe, (3) stellar winds, and (4) photo-ionization and photo-electric heating. Every star particle is treated as a single stellar population with known mass, age, and metallicity. The energy, momentum, mass, and metal returns from each stellar feedback processes are directly calculated from {\sc starburst99} \citep{leitherer.1999:sb99}.

\begin{figure}
\centering
\includegraphics[width=\linewidth]{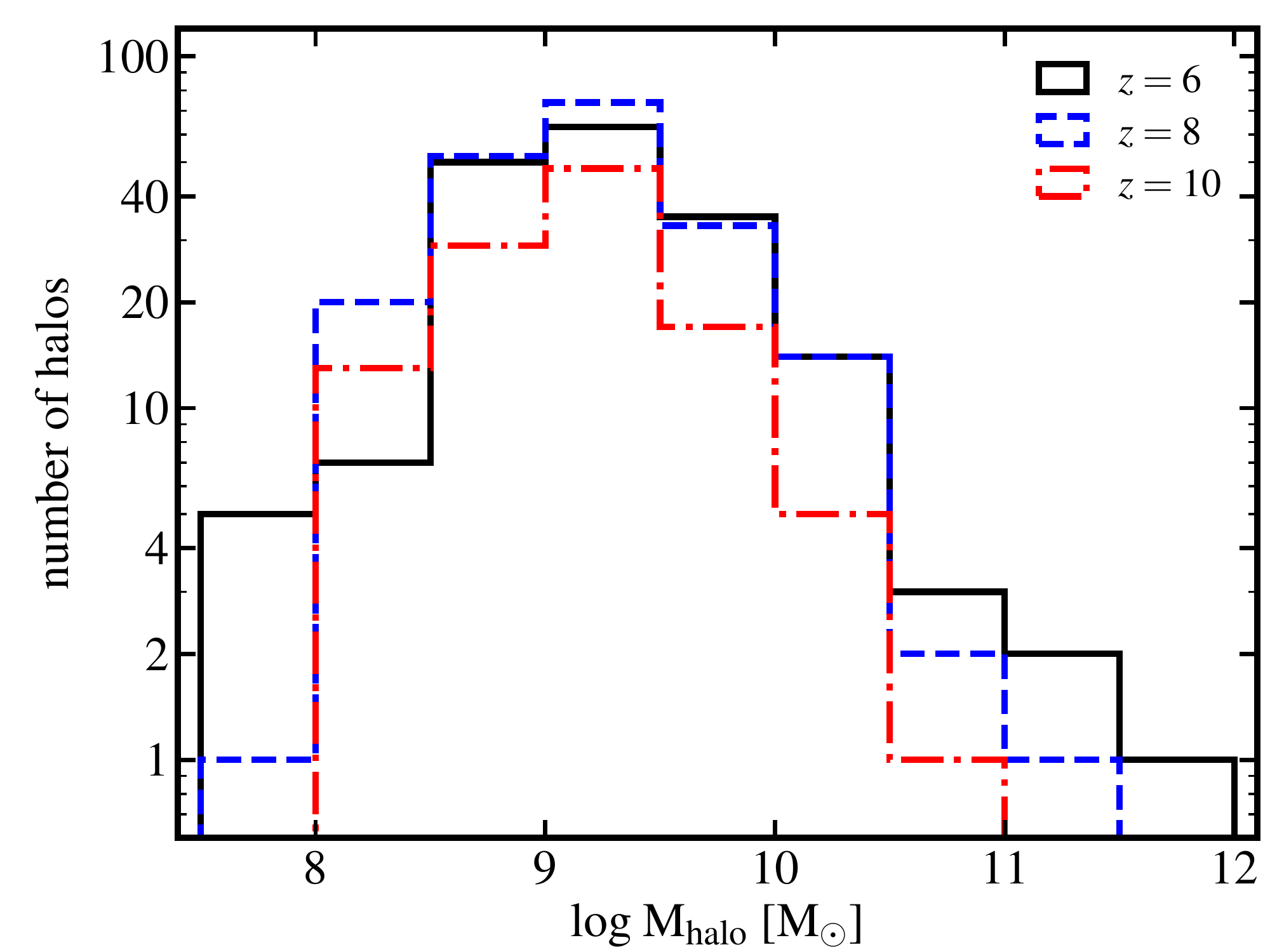}
\caption{\referee{Number of halos in our sample at $z=6$, 8, and 10.}}
\label{fig:nhalo}
\end{figure}

\begin{figure}
\centering
\includegraphics[width=\linewidth]{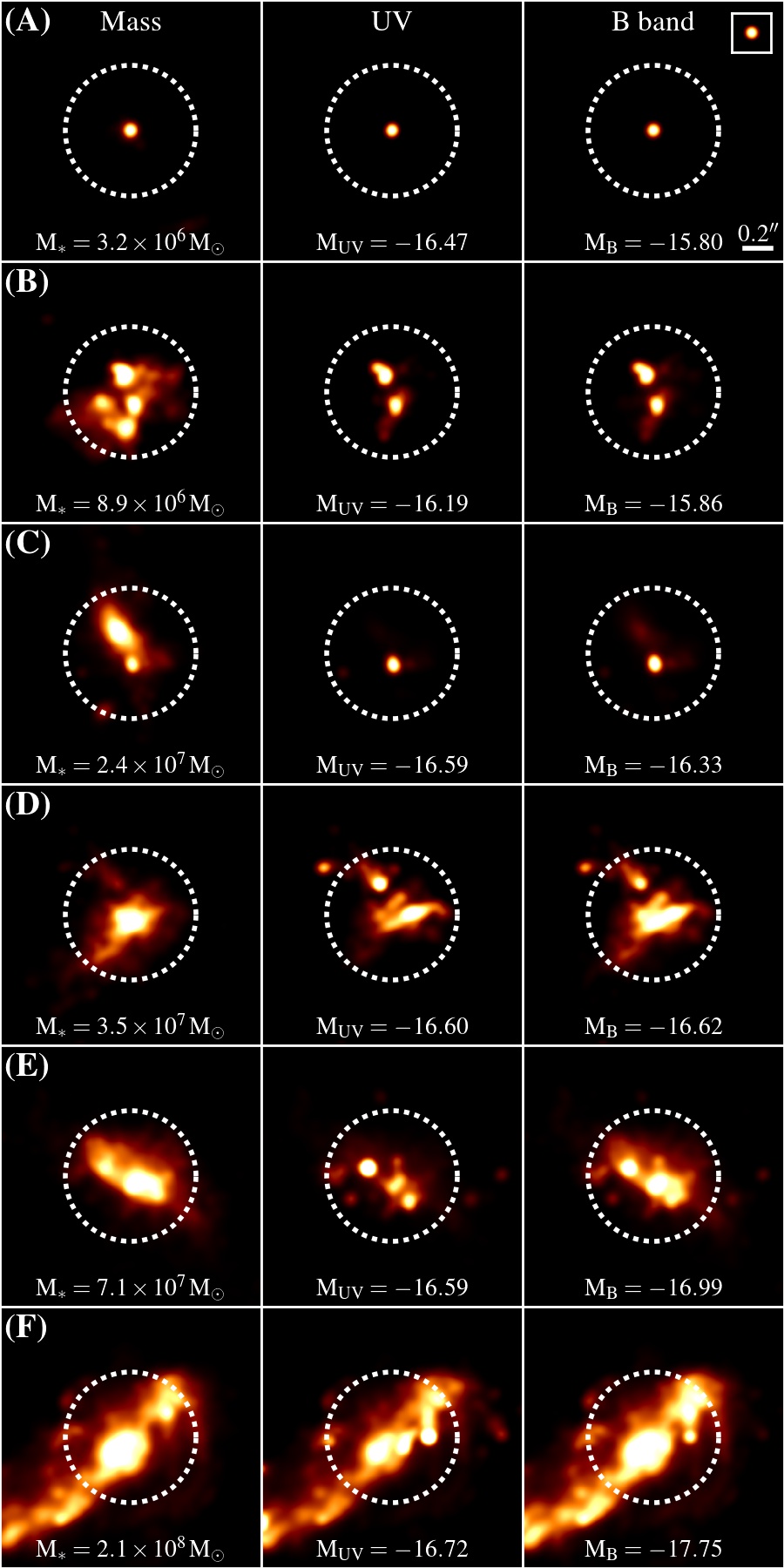}
\caption{Example images for six galaxies (A--F) at $z=6$ with rest-frame UV magnitude $\muv\sim -16.5$. The left column shows stellar surface density. The middle and right columns show unattenuated, noise-free rest-frame UV and rest-frame B-band surface brightness, respectively. All images are 2" (11.6\,kpc) along each side. Colors are in linear scale. These images are smoothed over a Gaussian kernel with 0.032" dispersion (10 pixels on the image) only for visualization purposes (otherwise the structures are too small to visualize on these images). The square in the top-right corner shows the appearance of a point source on these images for reference. The white dotted circles show the 1"-diameter aperture in which the sizes are measured. These galaxies span a wide range of stellar mass and B-band magnitude, and show a variety of morphologies. More massive galaxies appear to be larger than galaxies at lower masses. The UV images are largely dominated by bright, young stellar clumps, which do not necessarily trace the bulk of stars. The B-band light traces stellar mass better than the UV, but it can also be biased by the UV bright clumps. Galaxies tend to be more clumpy, more concentrated, and smaller in size from stellar mass to rest-frame optical to the UV.}
\label{fig:image}
\end{figure}

\subsection{Post-processing and size definition}
\label{sec:definition}
We use Amiga Halo Finder \citep[{\sc ahf};][]{knollmann.knebe.2009:ahf.code} to identify halos in our simulations. The halo mass ($\Mhalo$) and virial radius ($\Rvir$) are computed by {\sc ahf}, applying the redshift-dependent virial overdensity criterion from \citet{bryan.norman.1998:xray.cluster}. Each zoom-in simulation contains one central halo, which is the most massive halo of the zoom-in region. In this work, we also consider other less massive halos in the zoom-in region. We restrict our analysis to halos with zero contamination from low-resolution particles, which also having more than 100 star particles and $10^4$ total particles within the virial radius. \referee{In Figure \ref{fig:nhalo}, we show the number of halos in our simulated sample at $z=6$, 8, and 10.} At a given mass, our sample includes both central halos and less massive halos in the zoom-in regions, and include simulations run with different mass resolutions. In Appendix \ref{sec:append:res}, we show that our results converge reasonably well with resolution.

At a given redshift, we project all star particles inside the halo along a random direction onto a two-dimensional uniform grid. The pixel size of the grid is 0.0032 arcsec (3.2 mas), which equals to 1/10 of the pixel size of {\it JWST}'s Near-Infrared Camera (NIRCam) and corresponds to 10--20\,pc in the redshift range of our interest. Each star particle is smoothed over a cubic spline kernel with a smoothing length $h=1.5\,h_n$, where $h_n$ is its distance to the $n^{\rm th}$ nearest neighbor star particle. We adopt $n=5$ as our default value, but varying $n=5$--10 only makes small difference for a small fraction of our galaxies. We only consider intrinsic morphologies and sizes and do not include dust extinction in this work. The majority of galaxies (over 95\%) in our sample have intrinsic UV magnitude fainter than $\muv=-18$ (stellar mass $\Ms<10^8\,\Msun$). We find that dust attenuation has little effect on these low-mass, faint galaxies \citep[also see][]{ma.2017:fire.hiz.smf}, so most results in this paper are not affected by dust extinction.

We make projected images for stellar surface density and rest-frame UV (1500\,\AA) and rest-frame B-band (4300\,\AA) surface brightness. The rest-frame UV of galaxies at these redshifts falls in the short-wavelength channel of NIRCam (observed wavelengths 0.6--2.3\,$\mu$m, spatial resolution 0.032" per pixel), in F090W band for $z=5$ galaxies and F150W band for $z=10$ galaxies. The rest-frame B-band falls in NIRCam's long-wavelength channel (2.4--5\,$\mu$m, spatial resolution 0.065" per pixel), in F277W band for $z=5$ galaxies and F444W band for $z=10$ galaxies. The SED of each star particle is computed using the synthesis spectra predicted by the Binary Population and Spectral Synthesis (BPASS) models \citep[version 2.0;][]{eldridge.2008:bpass.model}\footnote{http://bpass.auckland.ac.nz}. We use the binary stellar population models in BPASS by default\footnote{\referee{Although we use {\sc starburst99} single-star models for stellar feedback in the simulations, BPASS binary models are not expected to strongly affect the feedback strength, since bolometric luminosities and supernovae rates are similar between these models \citep[see section 2.2 in][for more discussion]{ma.2017:fire.hiz.smf}. We prefer the binary models because they are able to reproduce the nebular emission line features in high-redshift galaxies \citep[e.g.][]{steidel.2016:stellar.nebular.spec,strom.2017:nebular.line.mosfire}.}}. In Figure \ref{fig:image}, we show example images for six galaxies labeled by A--F. Each panel is 2" (11.6\,kpc) on each side. Some galaxies and structures are so small that they only occupy very few pixels, so we further smooth the images using a two-dimensional Gaussian kernel with a dispersion equal to the size of 10 pixels (0.032") only for easier visualization here.

Most galaxies in our sample show clumpy, irregular morphologies that cannot be well described by a simple profile (see also figures 2 and 3 in \citealt{ma.2017:fire.hiz.smf}; cf. \citealt{jiang.2013:uv.morphology.z6}; \citealt{bowler.2017:bright.z7.galaxies}). Therefore, we adopt a non-parametric approach to define galaxy sizes, in a way similar to \citet{curtis-lake.2016:size.evolution} and \citet{ribeiro.2016:galaxy.size.evolution}. For every galaxy, we place a circular aperture of 1" in diameter, whose center is located by iteratively finding the B-band surface brightness-weighted center of all pixels within the 1"-aperture, as illustrated by the white dotted circles in Figure \ref{fig:image}. We visually inspect all galaxies to ensure the apertures are reasonably located. The same aperture is used for the size measurement in stellar mass, UV, and B-band luminosity for the same galaxy as follows. We sort the pixels enclosed in the 1"-diameter aperture in descending order of surface density or surface brightness, and find the number of `brightest' pixels that contribute 50\% of the total mass or luminosity within the 1" aperture. We calculate the area spanned by these pixels $S_{50}$ and define the `half-mass' or `half-light' radius as $R_{50}=\sqrt{S_{50}/\pi}$. We quote the galaxy stellar mass and luminosity as the total amount enclosed in the 1"-diameter aperture\footnote{\referee{We have checked that 1" aperture is sufficiently large for most galaxies in our sample, except for the few galaxies that are at a close encounter during a merger. In fact, galaxy F in Figure \ref{fig:image} is the mostly affected object, whose stellar mass and half-mass radius are underestimated by about $50\%$.}}.

One may also measure the half-mass or half-light radius alternatively by finding the radius that encloses half of the mass or light within some larger aperture. This is close to the commonly used algorithms in observations for size measurements, such as {\sc SExtractor} and {\sc galfit} \citep[e.g.][]{oesch.2010:hudf.morphology}, where concentric circular or ellipsoid apertures are usually assumed. However, this approach suffers from two main issues when applying to clumpy, irregular galaxies in our simulations. First, these galaxies do not have a well-defined center: one may use the position of intensity peak on the image or intensity-weighted center and get very different results. Second, for multi-clump systems (e.g. galaxies B, D, E, and F in the rest-frame UV, see Figure \ref{fig:image}), the size defined in this way in fact represents the spatial separation between the bright clumps. The non-parametric definition we use better reflects the physical size of individual clumps. For single-component objects, such as galaxy A in Figure \ref{fig:image} and well-ordered massive galaxies at intermediate and low redshifts, both definitions should give consistent results.

Nonetheless, we note that our size measurement depends on how we smooth the star particles. In general, using a larger smoothing length results in slightly larger galaxy sizes, but the difference is usually small for most of the galaxies. We refer the readers to Appendix \ref{sec:append} for details.

\begin{figure*}
\centering
\includegraphics[width=\linewidth]{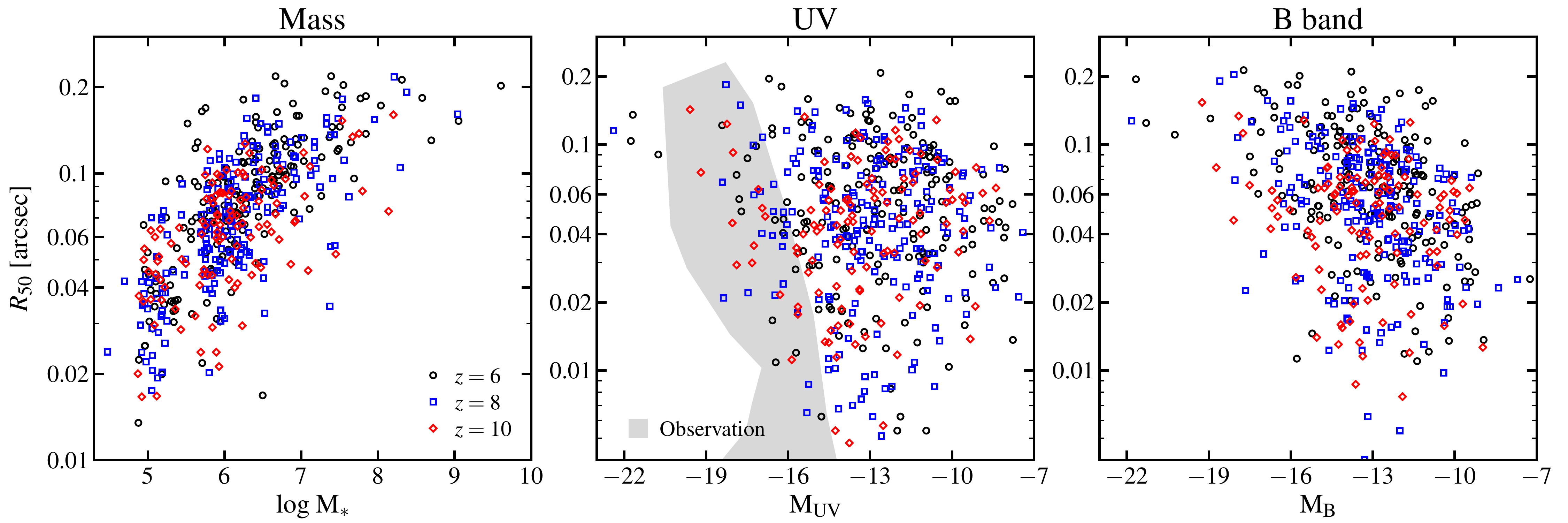}
\caption{Size--stellar mass relation (left) and size--luminosity relation in the rest-frame UV (middle) and rest-frame B band (right) at $z=6$ (black circles), 8 (blue squares) and 10 (red diamonds). At any redshift, the sizes of galaxies increase with stellar mass and/or luminosity, but all relations have considerable scatter. There is also a weak redshift evolution of galaxy sizes: high-redshift galaxies tend to have smaller sizes than low-redshift galaxies. The grey shaded region shows observational measurements from \citet{ono.2013:size.evolution.udf12}, \citet{kawamata.2015:hff.size.z6to8}, and \citet{bouwens.2017:small.galaxy.sizes}, adopted from the compilation in \citealt{bouwens.2017:small.galaxy.sizes} (see Section \ref{sec:szms} for a more detailed discussion).}
\label{fig:szms}
\end{figure*}

\begin{table*}
\caption{Best-fit parameters of galaxy size evolution.}
\begin{threeparttable}
\begin{tabular}{ccccccccc}
\hline
\multicolumn{3}{c}{Stellar mass} & \multicolumn{3}{c}{Rest-frame UV} & \multicolumn{3}{c}{Rest-frame B band} \\
\hline
$\log\Ms$ & $R_{50}~(z=5)$ & $m$ & $\rm M_{UV}$ & $R_{50}~(z=5)$ & $m$ & 
$\rm M_B$ & $R_{50}~(z=5)$ & $m$ \\
($\Msun$) & (kpc) & & (mag) & (kpc) & & (mag) & (kpc) & \\
\hline
$8$ & $1.09\pm0.04$ & $1.15\pm0.12$ & $-18$ & $0.76\pm0.08$ & $1.87\pm0.42$ & 
$-18$ & $0.99\pm0.07$ & $1.53\pm0.25$ \\
$7$ & $0.91\pm0.02$ & $1.49\pm0.09$ & $-16$ & $0.55\pm0.05$ & $1.78\pm0.32$ & 
$-16$ & $0.74\pm0.04$ & $1.63\pm0.19$ \\
$6$ & $0.52\pm0.02$ & $0.87\pm0.09$ & $-14$ & $0.49\pm0.01$ & $1.78\pm0.06$ & 
$-14$ & $0.66\pm0.02$ & $1.92\pm0.11$ \\
$5$ & $0.29\pm0.01$ & $0.84\pm0.11$ & $-12$ & $0.45\pm0.02$ & $1.00\pm0.15$ & 
$-12$ & $0.46\pm0.01$ & $1.05\pm0.09$ \\
\hline
\end{tabular}
\begin{tablenotes}
\item Note: The size evolution of galaxies at a given stellar mass or magnitude is described as $R_{50}\sim(1+z)^{-m}$ (see Section \ref{sec:evolution} for details). $R_{50}~(z=5)$ gives the best-fit normalization at $z=5$.
\end{tablenotes}
\end{threeparttable}
\label{tbl:sz}
\end{table*}

\section{Results}
\label{sec:results}

\subsection{Qualitative behaviors: an overview}
\label{sec:overview}
In Figure \ref{fig:image}, we show projected images of stellar mass (left), and noise-free rest-frame UV (middle) and rest-frame B-band (right) luminosity for six galaxies at $z=6$. These galaxies are selected to have similar UV magnitudes around $\muv\sim-16.5$, in increasing order of stellar mass from the top to the bottom. The images are smoothed for visualization purposes in Figure \ref{fig:image}. Our default size measurements are performed using the original images. The colors represent surface density or surface brightness in linear scale and they saturate at a level such that pixels above it contribute 10\% of the total intensity on the image. The square in the top-right corner shows the appearance of a point source on these images.

Despite all galaxies having $\muv\sim-16.5$, they span two orders of magnitude in stellar mass from $\Ms=3\times10^6$--$2\times10^8\,\Msun$. They show a wide range of morphologies in their surface density maps: galaxy A is compact; galaxies C, E, and F all have a small companion that is close (within 0.2") to the main galaxy; galaxy B is made of several clumps of comparable sizes. High-mass galaxies are generally larger than low-mass galaxies in all bands.

The UV images of these galaxies are largely dominated by one or several bright clumps, where the stellar populations are relatively young (10\,Myr on average). Most of the UV clumps are intrinsically small and appear like point sources on these images. More importantly, these bright clumps do not necessarily trace the bulk of stars. In galaxy C, for example, the UV bright clump is associated with the small companion, while the larger, more massive main galaxy is very faint in the UV. In galaxy D, there are two dominant clumps in the UV image: the smaller one to the upper-left to the galaxy is not associated with any prominent stellar structure; the larger one also has a small spatial offset to the right of the stellar surface density peak. We visually inspected every galaxy in our sample and found this phenomenon to be very common in our simulated galaxies (e.g. galaxies E and F). This is consistent with the off-center star-forming UV clumps observed in intermediate-redshift galaxies \citep[$z\sim0.5$--2.5, e.g.][]{wuyts.2012:candels.clumps}, which can contribute a large fraction of star formation but only a small fraction of stellar mass. These clumps are either small satellite galaxies or stars formed in individual star-forming regions (see Section \ref{sec:clump} for more discussion)\footnote{\referee{Each system enclosed by an 1" aperture is counted as one galaxy in this paper, even if it contains multiple clumps. We consider that the clumps are sufficiently close to be associated as one galaxy.}}.

In contrast, the B-band light traces stellar mass better than the UV, although it can also be biased by the young, bright stellar clumps in some circumstances. In galaxy E, the UV bright clump associated with the companion upper-left to the central galaxy is also bright in B-band. The central galaxy also appears bright in the B-band, but it is much fainter in the UV, due to an relatively older stellar population than the UV-bright stellar clump. In galaxy C, the B-band image is dominated by the only bright clump; the main galaxy, however, is faint in the B band, because its stellar population is much older.

In general, galaxy size increases with increasing stellar mass or luminosity, following the size--mass or size--luminosity relation. From stellar mass to rest-frame optical to the UV, galaxies tend to be more clumpy, more concentrated, and smaller in size. Moreover, there is a broad distribution of galaxy UV size at fixed UV magnitude. Galaxies A--C have intrinsically small UV sizes, because nearly all of the UV light is emitted by the bright clumps. In contrast, in galaxies D--F, the more extended, low surface brightness pixels contribute a large fraction of the total UV luminosity, so they have larger half-light radius in the UV. However, we caution that low surface brightness regions may fall below the detection limit of a given observational campaign, so the observed half-light radius is consequently smaller if the imaging decreases in depth (Section \ref{sec:szmu}). A better way to compare our simulations with observations is to add the background noise of an observing campaign and process the simulations with an identical pipeline for size measurement on the mock images. This is beyond the scope of the current paper, but it is worth exploring in the future.

\subsection{Size--mass and size--luminosity relations}
\label{sec:szms}
In Figure \ref{fig:szms}, we show the galaxy size--mass relation (left) and size--luminosity relation in the UV (middle) and B band (right) for our simulated sample. The points represent galaxies at $z=6$ (black circles), 8 (blue squares), and 10 (red diamonds). We follow \citet{bouwens.2017:small.galaxy.sizes} and express the sizes in arcsec in this section.  At any redshift, there is a correlation between galaxy size and stellar mass and/or luminosity with considerable scatter. At $\Ms<10^8\,\Msun$, the half-mass radius spans a factor of 3 (0.5\,dex) at fixed stellar mass. The scatter is likely driven by several different effects, including halo-to-halo variance, and dynamical effects connected to mergers and strong stellar feedback \citep{elbadry.2016:fire.migration}, and mergers. The size--luminosity relations show larger scatter: at $\rm M_{UV}>-18$ and $\rm M_B>-18$, the half-light radii spans nearly one dex at fixed magnitude. Most simulated galaxies have half-light radii within 0.01--0.2", while some galaxies have even smaller half-light radii down to 0.005". In contrast, very few galaxies have half-mass radii smaller than 0.02", suggesting that galaxies with extremely small UV sizes should be larger in terms of stellar mass. This is because the bright clumps that dominate the light in these galaxies are very concentrated. At the more massive/brighter end, our simulations do not contain sufficient number of galaxies for a robust estimate of the scatter. In addition, there is a weak redshift evolution of galaxy sizes: the median angular size of galaxies decreases by 25\% (physical size by a factor of two) from $z=6$ to $z=10$ at a fixed stellar mass and/or magnitude. This is much smaller than the intrinsic scatter of the size--mass and size--luminosity relations (see Section \ref{sec:evolution} for quantitative results).

The grey shaded region in Figure \ref{fig:szms} shows the observational data taken from the compilation in \citet{bouwens.2017:small.galaxy.sizes} (also including the data from \citealt{ono.2013:size.evolution.udf12} and \citealt{kawamata.2015:hff.size.z6to8}). \citet{kawamata.2015:hff.size.z6to8} measured the sizes of 31 lensed galaxies at $z=6$--8 in the Abell 2744 cluster field from the HFF. The half-light radii of galaxies at $\rm M_{UV}\sim-19.5$ in their $z\sim6$--7 sample range from 0.1--1\,kpc, corresponding to 0.02--0.2" at these redshifts. Similarly, \citet{bouwens.2017:small.galaxy.sizes} also found a similar range of half-light radius from 0.02--0.2" for galaxies with $\rm M_{UV}\sim-18.5$ at $z\sim6$ in a more complete HFF sample. \citet{ono.2013:size.evolution.udf12} found that $z\sim7$--8 galaxies of $\rm M_{UV}\sim-19$ in the HUDF also have half-light radii from 0.02--0.2" with a median of about 0.06". Brighter galaxies at $\rm M_{UV}\sim-21$ in the HUDF have half-light radii about 0.15" at $z\sim5$--8 \citep[e.g.][]{bouwens.2004:hudf.size.evolution,oesch.2010:hudf.morphology}. The most massive galaxies in our sample broadly agree with these measurements.

So far, only a small sample of galaxies fainter than $\rm M_{UV}\sim-18$ from the HFF have size measurements \citep{kawamata.2015:hff.size.z6to8,bouwens.2017:small.galaxy.sizes}. These galaxies are reported to have very small intrinsic sizes from 0.01--0.06", and a small fraction of them even have half-light radii down to 0.001". Some of our simulated galaxies fall in the observed range, but our sample also contains a large number of galaxies that have much larger sizes (they tend to lie above the grey shaded region at a given magnitude in Figure \ref{fig:szms}). We speculate two possible observational biases/uncertainties that may lead to such discrepancies. 
First, at fixed magnitude, larger galaxies tend to have lower surface brightness, so they are more likely to be missed in the observed sample. Second, for clumpy galaxies, one may only pick up the brightest clumps and thus the sizes are underestimated. Therefore, observations are likely biased toward intrinsically small galaxies and/or star-forming clumps \citep[e.g.][]{vanzella.2017:globular.formation}. In Section \ref{sec:szmu}, we will explicitly show the effect of limited surface brightness sensitivity on the observed galaxy sizes. Future deep observations on a few candidate clumpy galaxies with {\it JWST} can test our predictions. On the other hand, we note that our size measurements are different from those commonly used in observations (see Section \ref{sec:definition} for a detailed discussion), which further complicates the comparison. It would be interesting to carry out more detailed comparisons with specific observational campaigns to understand the discrepancies.

\begin{figure}
\centering
\includegraphics[width=\linewidth]{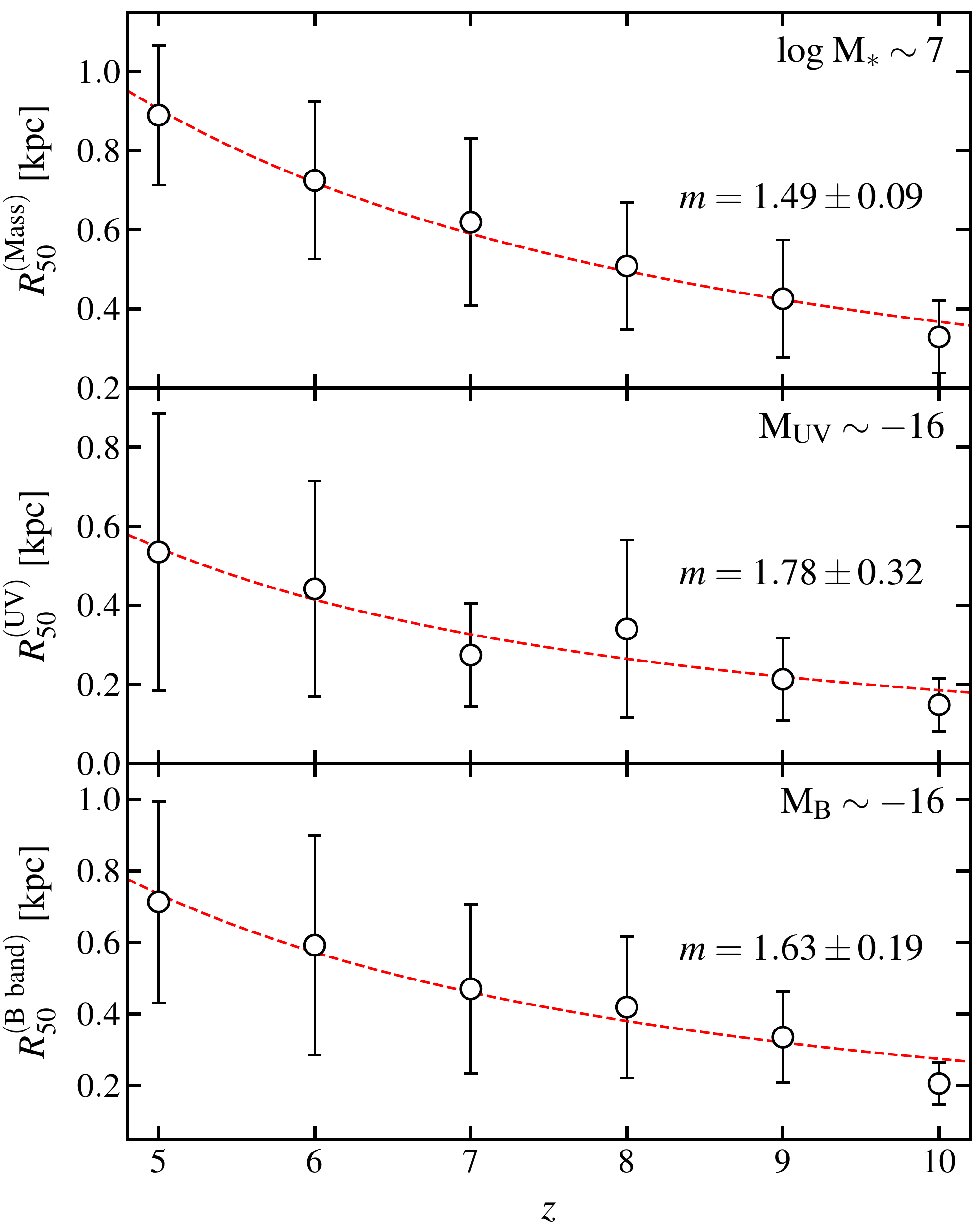}
\caption{Size evolution of the simulated galaxies at $\Ms\sim10^7\,\Msun$ (top), $\rm M_{UV}\sim-16$ (middle), and $\rm M_B\sim-16$ (bottom). Points with errorbars show the mean and 1$\sigma$ (16--84\%) distribution of physical half-light radii (in kpc) at $z=5$--10. The red lines show the best-fit evolution following $R_{50}\sim(1+z)^{-m}$. The best-fit parameters are also listed in Table \ref{tbl:sz}.}
\label{fig:sz}
\end{figure}

\begin{figure}
\centering
\includegraphics[width=\linewidth]{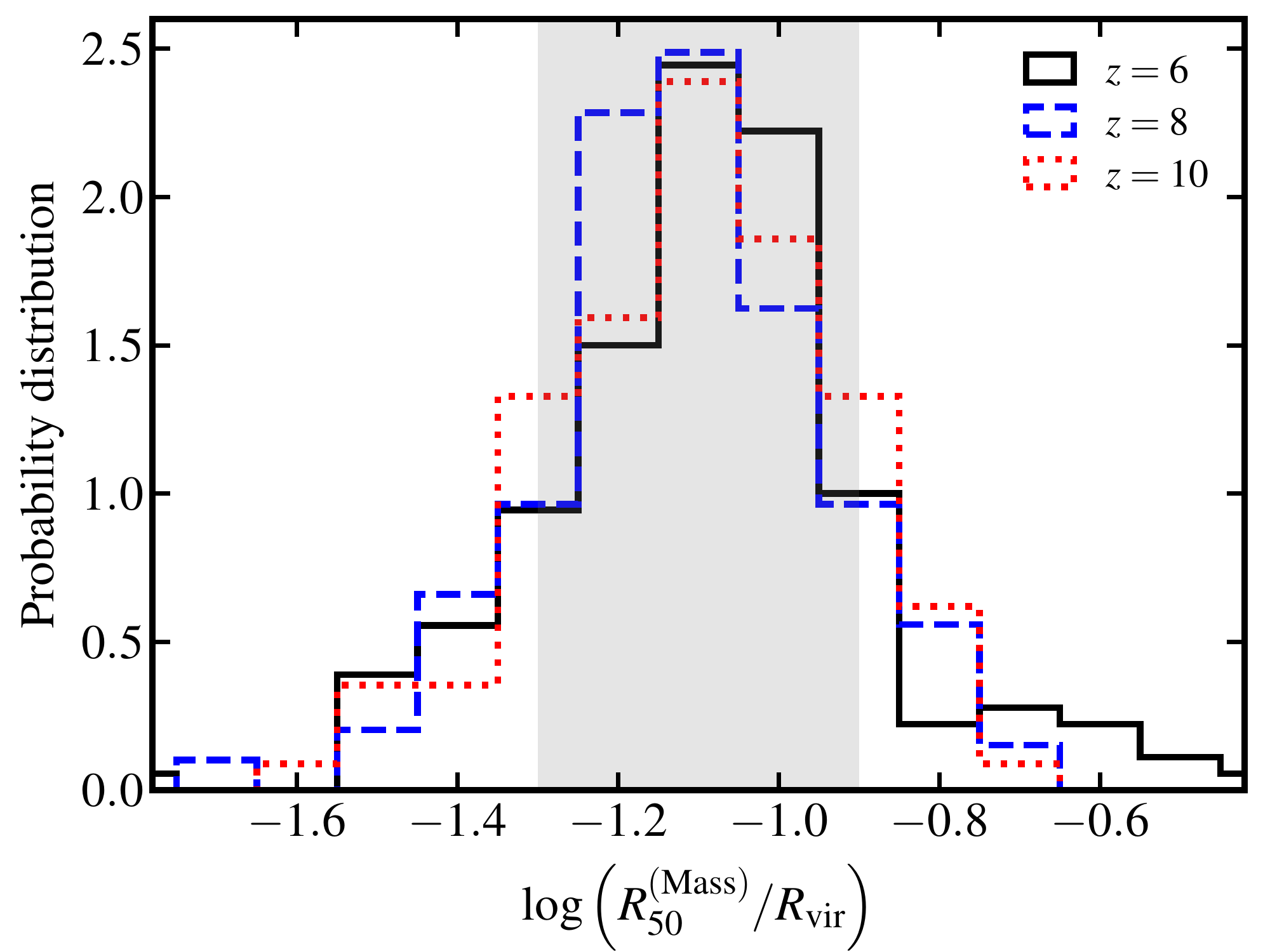}
\caption{\referee{Probability distribution (normalized) of galaxy half-mass to halo virial radius ratio for the simulated sample at $z=6$, 8, and 10. $R_{50}^{\rm (Mass)}/R_{\rm vir}$ has a median of 8\% and $1\sigma$ range of 5--12\% (the shaded region). The distribution does not significantly evolve at these redshifts.}}
\label{fig:dist}
\end{figure}

\begin{figure*}
\centering
\includegraphics[width=\linewidth]{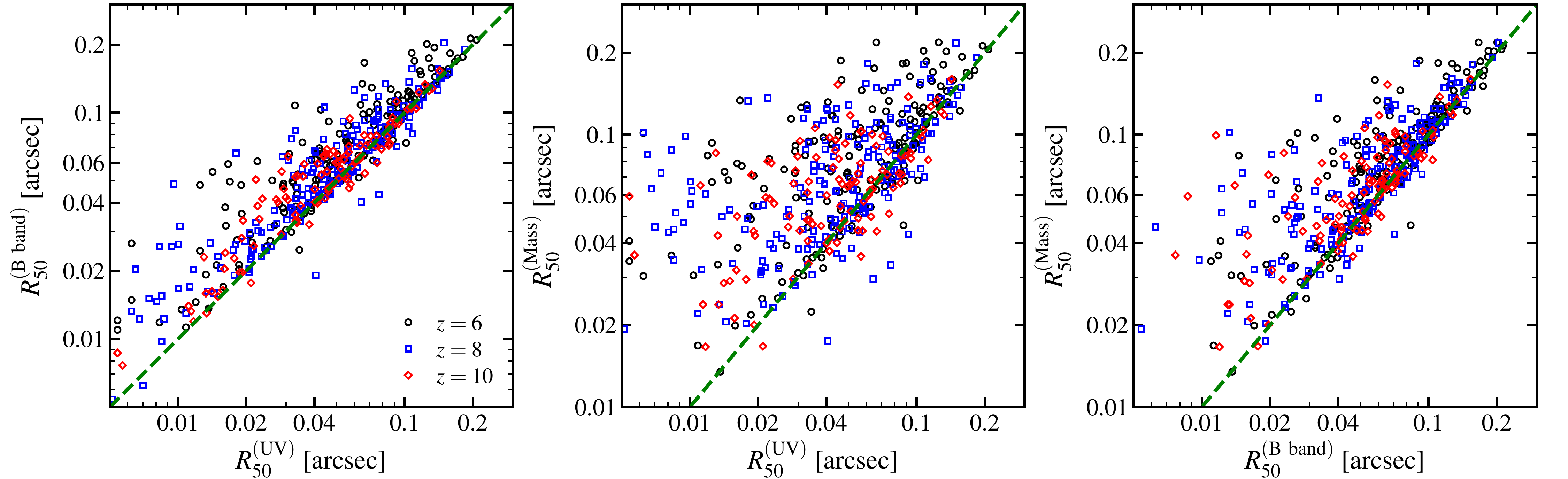}
\caption{The comparison of galaxy sizes measured in stellar mass, rest-frame UV, and rest-frame B band. The half-mass (light) radii systematically decrease from stellar mass to B band to the UV, in line with a more concentrated morphology in this sequence. The $R_{50}^{\rm (Mass)}$--$R_{50}^{\rm (UV)}$ relation shows larger scatter than the $R_{50}^{\rm (Mass)}$--$R_{50}^{\rm (B~band)}$ relation, suggesting the UV light is a relatively worse tracer of the stellar mass distribution. Galaxies with small UV sizes are also small in B band, but they usually have much larger half-mass radii, because the B-band light is biased by the bright clumps in these galaxies.}
\label{fig:szsz}
\end{figure*}

\subsection{Size evolution}
\label{sec:evolution}
In this section, we quantify the redshift evolution of galaxy sizes using the simulated sample at $z=5$--10. At each redshift, we bin our data in stellar mass every $\Delta\log\Ms=1$ and/or in magnitude every 2\,mag. In each bin, we calculate the mean and $1\sigma$ distribution (14--86 percentile) of galaxy half-mass and/or half-light radii. In Figure \ref{fig:sz}, we show the results at $\Ms\sim10^7\,\Msun$ (top), $\rm M_{UV}\sim-16$ (middle), and $\rm M_B\sim-16$ (bottom) (same stellar mass/magnitude at all redshifts). Note that we show the physical sizes (in kpc) instead of angular sizes (in arcsec). We fit the evolution trend with a functional form $R_{50}\sim(1+z)^{-m}$. The red dashed lines in Figure \ref{fig:sz} show the best-fit results at these bins. In Table \ref{tbl:sz}, we list all the best-fit parameters from $\Ms\sim10^5$--$10^{8}\,\Msun$, $-18<\rm M_{UV}<-12$, and $-18<\rm M_B<-12$. It is worth noting that the evolution of the physical sizes has power-law index $m\sim1$--2, which is steeper than the redshift dependence of the angular diameter distance [$D_A\sim(1+z)^{2/3}$]. This indicates that the angular sizes of galaxies also decrease with redshift, as shown in Figure \ref{fig:szms}.

There are some observational constraints on the size evolution. Various authors have reported $m\sim1$--1.5 for galaxies brighter than $\rm M_{UV}<-19$ across $z\sim0$--8 \citep[e.g.][]{bouwens.2004:hudf.size.evolution,oesch.2010:hudf.morphology,kawamata.2015:hff.size.z6to8,shibuya.2015:legacy.size.evolution}. Our results show broad agreement with these constraints (within $1\sigma$ for most mass and magnitude bins), although we mainly study galaxies at lower masses and luminosities than the observed sample, and only focus on $z\geq5$. We note that the best-fit value of $m$ is sensitive to the data and their uncertainties. For several stellar mass and magnitude bins, our sample only contains a small number of galaxies at some redshift. Stochastic effects in bins with small numbers of objects can strongly affect the results of fitting in those bins.

\referee{In Figure \ref{fig:dist}, we show the distribution of the ratio of galaxy half-mass radius to halo virial radius $R_{50}^{\rm (Mass)}/R_{\rm vir}$ for our simulated sample at $z=6$, 8, and 10.} We find that for the entire sample, this ratio has a median of 8\% and $1\sigma$ range from 5--12\% (the shaded region in Figure \ref{fig:dist}). The median and dispersion do not strongly evolve with redshift at $z=5$--10. This is consistent with the fact that $R_{50}^{\rm (Mass)}\sim(1+z)^{-m}$ with $m\sim1$ at these masses (see Table \ref{tbl:sz}), given a non-evolving stellar mass--halo mass relation at these redshifts \citep{ma.2017:fire.hiz.smf} and $\Rvir\sim(1+z)^{-1}$ at a fixed halo mass (the virial overdensity is nearly a constant at these redshift; see \citealt{bryan.norman.1998:xray.cluster}). For the few more massive galaxies in our sample ($\Ms>10^8\,\Msun$), this ratio is smaller, mostly at 1--5\%: this is comparable to observational measurements for galaxies at similar masses \citep[$\sim3\%$, e.g.][]{kawamata.2015:hff.size.z6to8,shibuya.2015:legacy.size.evolution}. Our simulations thus predict that at these redshifts, the stellar-to-halo size ratio (as defined above) is larger for low-mass galaxies, where there are no observational constraints so far.

Our results at $z\geq5$ should not be extrapolated to lower redshifts. Recently, \citet{fitts.2017:field.dwarf.galaxy} presented a suite of cosmological zoom-in simulations of isolated dwarf galaxies run to $z=0$ using the same FIRE-2 code. All of their galaxies are hosted in halos of $\Mhalo\sim10^{10}\,\Msun$ at $z=0$. Several galaxies in their sample have stellar mass $\Ms\sim10^7\,\Msun$: these are all early-forming galaxies with half-mass radii around 1\,kpc. This is very close to our $z=5$ galaxy sizes at the same stellar mass, likely due to the fact that the early-forming galaxies in \citet{fitts.2017:field.dwarf.galaxy} do not grow significantly at later times. Although this is a biased sample, and $\Ms\sim10^7\,\Msun$ galaxies may have a broad distribution of half-mass radius at $z=0$, this suggests that the stellar-to-halo size ratio may be much at lower redshifts for low-mass galaxies (since the virial radius increases with decreasing redshift at fixed mass). This could be due, for example, to less efficient halo gas accretion at later times. For more massive galaxies, our simulations show that the stellar-to-halo size ratio at $z\geq5$ is already comparable to that at $z\sim0$ ($\sim2\%$), so it may not evolve strongly at later times \citep[e.g.][]{shibuya.2015:legacy.size.evolution}. A mass-dependent evolution of the stellar-to-halo size ratio is consistent with recent analysis for $z\lesssim3$ galaxies \citep[e.g.][]{somerville.2018:size.halo.evolution}. The galaxy size and morphology evolution down to $z\sim0$ will be studied in details in a separate paper (Schmitz et al., in preparation).

\subsection{Galaxy sizes at different bands}
\label{sec:szsz}
In Section \ref{sec:overview}, we show examples of our simulated galaxies to illustrate that galaxies tend to be more concentrated and smaller from stellar mass to B band to the UV. The UV light is dominated by small, bright, young stellar clumps, while the B-band morphology is determined by both bright clumps and more broadly distributed stars. In Figure \ref{fig:szsz}, we compare the half-mass (light) radii measured in one quantity against another. The green dashed lines show the $y=x$ relation. All three sizes correlate with each other, but $R_{50}^{\rm (B~band)}$ is systematically larger than $R_{50}^{\rm (UV)}$, and the $R_{50}^{\rm (Mass)}$ is larger than both $R_{50}^{\rm (UV)}$ and $R_{50}^{\rm (B~band)}$. We also check the Gini coefficient \citep[e.g.][]{lotz.2004:non.parametric.morph}, which is a parameter between 0 and 1 that describes the concentration of galaxy morphology (1 being the most concentrated). We find that the Gini coefficient increases from stellar mass to B band to the UV, in line with the decreasing galaxy sizes in the sequence.

The correlation between $R_{50}^{\rm (Mass)}$ and $R_{50}^{\rm (UV)}$ has a larger scatter than that between $R_{50}^{\rm (Mass)}$ and $R_{50}^{\rm (B~band)}$, indicating that rest-frame UV light is a relatively worse tracer of stellar mass than the B band. Furthermore, galaxies with intrinsically small UV sizes (below 0.01") mostly have small B-band sizes (below 0.02") as well, although these galaxies usually have relatively large half-mass radii (0.04--0.1"). This is because the B-band light is also biased by the small, bright clumps with high light-to-mass ratios in these galaxies (e.g. galaxy C in Figure \ref{fig:image}).

\begin{figure}
\centering
\includegraphics[width=\linewidth]{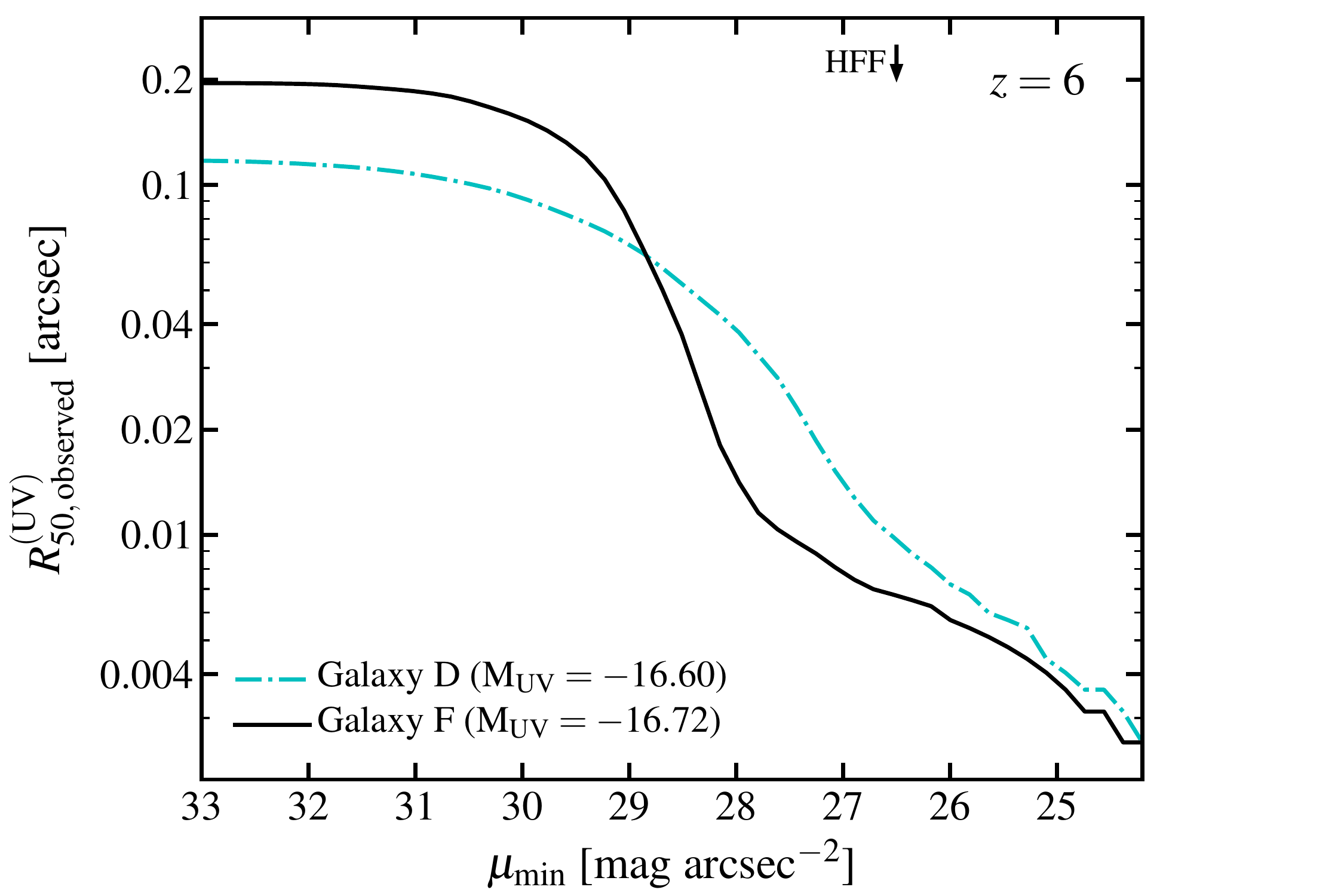} \\
\vspace{0.4cm}
\includegraphics[width=\linewidth]{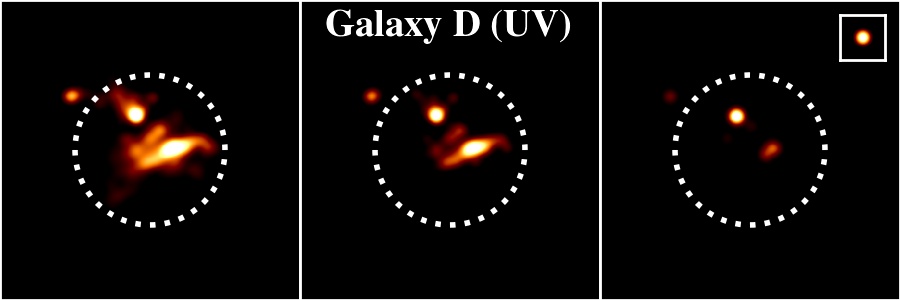} \\
\includegraphics[width=\linewidth]{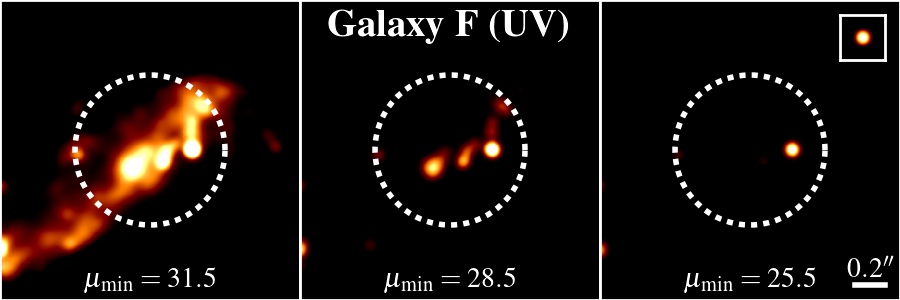} 
\caption{{\em Top:} Galaxy UV half-light radii measured assuming different surface brightness detection limits for galaxies D and F from Figure \ref{fig:image}. The `observed' size increases with the depth of imaging. {\em Bottom:} Appearance of galaxies D and F (Figure \ref{fig:image}) at different rest-frame UV surface brightness limits. At a detection limit of $\mu_{\rm min}=25.5$\,mag\,arcsec$^{-2}$, the galaxies appear as point sources, and only the brightest clump is dominant.}
\label{fig:szmu}
\end{figure}

\begin{figure}
\centering
\includegraphics[width=\linewidth]{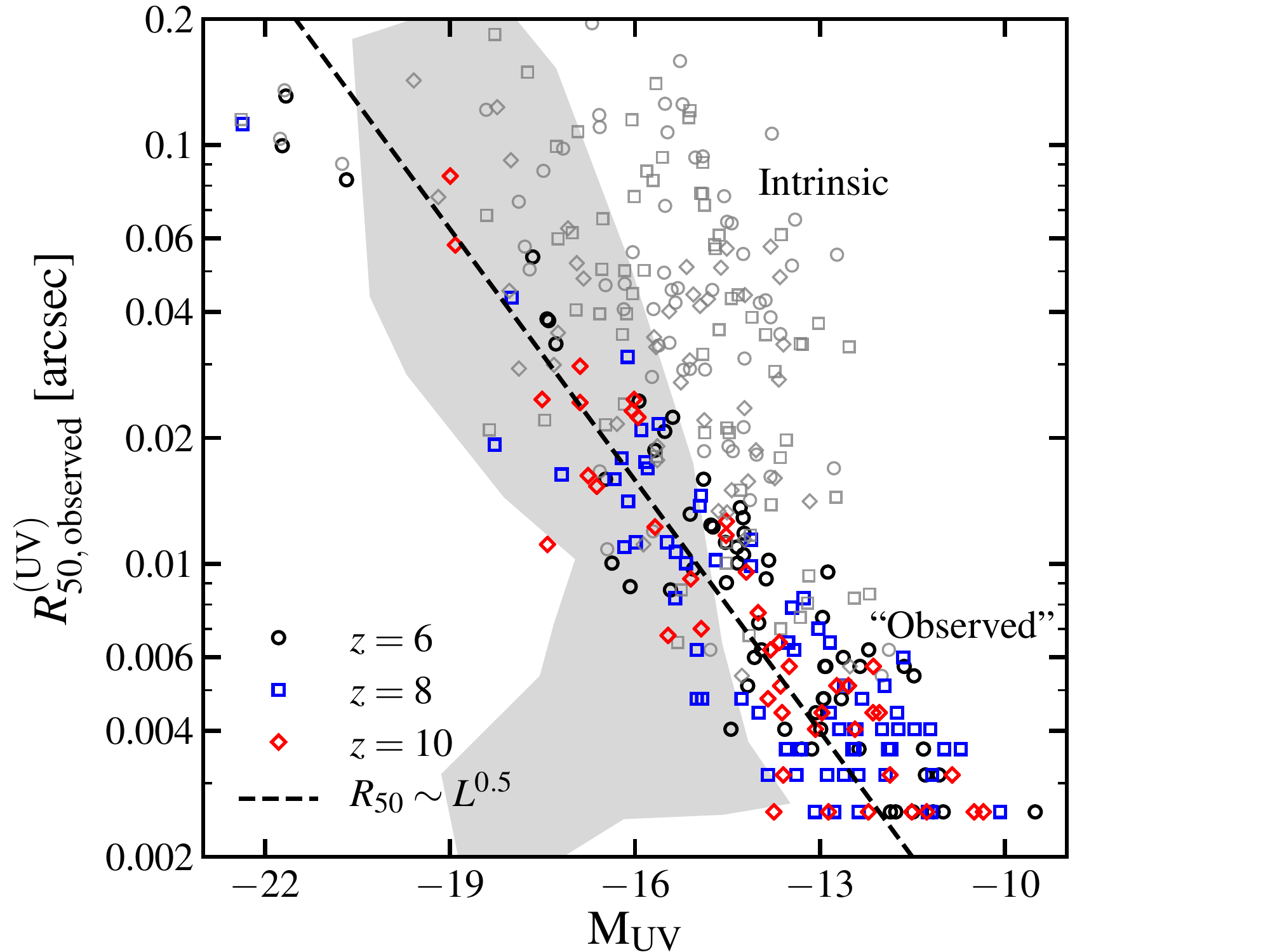} 
\caption{The `observed' rest-frame UV size--luminosity relation for our simulated sample after mimicking the effect of the HFF surface brightness detection limit at $\mu_{\rm min}\sim26.5$\,mag\,arcsec$^{-2}$. The grey shaded region represents the observational data as in Figure \ref{fig:szms}. The grey points show the intrinsic size--luminosity relation for the same galaxies (non-detectable galaxies are not shown). Most galaxies appear fainter and show much smaller `observed' sizes. The black dashed line shows the $R_{50}\sim L^{0.5}$ scaling as suggested in \citet{bouwens.2017:small.galaxy.sizes} for HFF galaxies; however, such scaling is expected due to the selection effect of a surface brightness-limited sample ($L/R^2$ is constant).}
\label{fig:hff}
\end{figure}

\section{Discussion}
\label{sec:discussion}

\subsection{How do surface brightness limits affect observed galaxy sizes?}
\label{sec:szmu}
In some galaxies, a large fraction of the total UV luminosity is contributed in low surface brightness, diffuse light (pixels). These regions are dominated by relatively older stars (10--100\,Myr) with lower light-to-mass ratios than those in the young, bright clumps (e.g. galaxies D--F in Figure \ref{fig:image}). For a specific observing campaign, there is a surface brightness limit below which the signal-to-noise ratio is too low to be detectable: this can have a significant effect on the observed morphologies and size measurements of clump-dominated galaxies. In the top panel of Figure \ref{fig:szmu}, we illustrate this effect using example galaxies D and F from Figure \ref{fig:image} at $z=6$. We show the rest-frame UV half-light radii measured for the same galaxies as a function of surface brightness limit (assuming pixels below such limit have zero flux). The effect can be dramatic in some circumstances: for galaxy F, the `observed' half-light radius decreases by over an order of magnitude (from 0.1" to 0.01") if the surface brightness depth drops from 29 to 28\,mag\,arcsec$^{-2}$. In the bottom panel of Figure \ref{fig:szmu}, we further show the rest-frame UV images of galaxies D and F at three surface brightness limits. From $\mu_{\rm min}=31.5$ to 28.5\,mag\,arcsec$^{-2}$, most of the low-surface-brightness regions become `invisible', and the galaxy is dominated by a few clumps in the UV. Once the detection limit further drops to $\mu_{\rm min}=25.5$\,mag\,arcsec$^{-2}$, only the intrinsically small, brightest clump is dominant in these galaxies as a point source.

Now we discuss the implications of this effect on the size--luminosity relation and extremely small sizes measured for galaxies in the HFF. The typical $5\sigma$ point-source detection limit in the rest-frame UV of $z=5$--10 galaxies is $\sim28.7$--29.1\,mag within a 0.4"-diameter aperture \citep{coe.2015:hubble.frontier.field}. This corresponds to a surface brightness limit about $\mu_{\rm min}\sim26.5$\,mag\,arcsec$^{-2}$ for extended sources if we demand the same signal-to-noise ratio within the same aperture. As a proof of concept, we perform a simple experiment on our simulated galaxies to mimic the HFF detection limit: we zero out all pixels below 26.5\,mag\,arcsec$^{-2}$ and re-measure the luminosities and sizes. We find that {\em all} galaxies intrinsically brighter than $\muv<-13$ are still detectable, but their `observed' luminosities and sizes become smaller. {\em No} galaxies intrinsically fainter than $\muv>-12$ are detectable. Approximately, the fraction of light {\em lost} due to such surface brightness cut is a linear function of intrinsic UV magnitude, from zero at $\muv=-22$ to unity at $\muv=-12$. In Figure \ref{fig:hff}, we show the `observed' size--luminosity relation in the rest-frame UV for our simulated sample. The intrinsic size--luminosity relation for the same galaxies is shown by grey points for reference (non-detectable galaxies are not shown). Most galaxies appear fainter and have much smaller `observed' sizes.  When taking into account surface brightness limits, our simulations broadly follow a $R_{50}\sim L^{0.5}$ relation (the black dashed line), as suggested in \citet{bouwens.2017:small.galaxy.sizes} for HFF galaxies, but this trend is affected by the $L/R^2\sim{\rm constant}$ selection for a given surface brightness limit. Nonetheless, our simple experiment is by no means a one-to-one comparison with HFF observations. Ideally, one should post-process the high-resolution images of simulated galaxies with gravitational lensing, convolve them with HST PSF, add comparable background noise, run identical source finder, and measure the luminosities and sizes using the same method \citep[e.g.][]{price.2017:size.recovery.fire}. This is beyond the scope of this paper, but is worth future exploration in parallel with JWST deep surveys.

\subsection{Implications for the observed (faint-end) galaxy UV luminosity functions}
\label{sec:hff}
Current observational constraints on the $z\gtrsim6$ galaxy UV luminosity functions fainter than $\rm M_{UV}\sim-17$ come from the HFF program, which takes advantages of foreground galaxy clusters to detect strongly gravitationally lensed high-redshift galaxies. Our results in this paper have two important implications for these observations. First, our simulations show a broad distribution of galaxy sizes at fixed UV magnitude. This affects the estimated completeness correction for the observed sample: if there are more galaxies that have large sizes than expected (they cannot be detected due to low surface brightness), their number densities may be underestimated \citep{bouwens.2017:small.galaxy.sizes}. Second, some galaxies are dominated by a few small, bright clumps in the rest-frame UV, so they can be mis-identified as several fainter galaxies. If this is the case, the UV luminosity function can be underestimated at intermediate magnitudes, but overestimated at fainter magnitudes. It is interesting that some faint-end UV luminosity functions derived from HFF samples show a small discontinuity at the magnitude where this effect is likely to become important \citep[although it may also be caused by other effects, e.g.][]{bouwens.2017:lensing.uncertainty,livermore.2017:faint.galaxies}. A more quantitative analysis of the observational biases is worth future investigation.

\begin{figure}
\centering
\includegraphics[width=\linewidth]{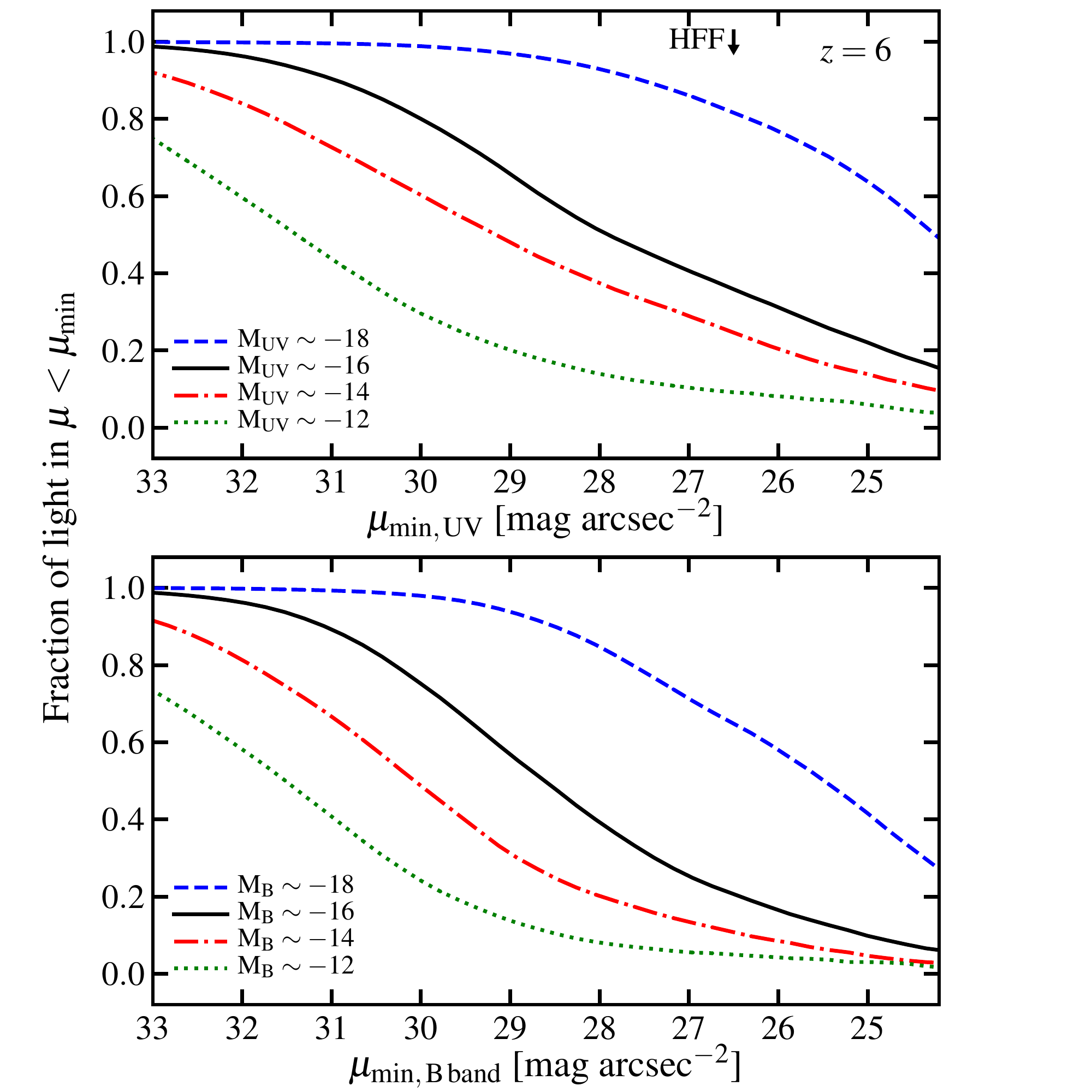} 
\caption{The fraction of light in pixels brighter than $\mu<\mu_{\rm min}$ as a function of $\mu_{\rm min}$, averaged over the simulated galaxies at a given intrinsic magnitude (total luminosity) in the rest-frame UV (top) and B band (bottom) at $z=6$. These results provide predictions on what depths one needs to reach, to target galaxies at a certain magnitude.}
\label{fig:flight}
\end{figure}

\subsection{What fraction of light come from low surface brightness regions?}
\label{sec:flight}
In this section, we attempt to address the following question: for a given surface brightness limit, what fraction of a galaxy's light will be detected or missed? This is useful for planning future JWST deep surveys or follow-up deep imaging and understanding the completeness of an observed sample. In Figure \ref{fig:flight}, we show the fraction of light in pixels brighter than $\mu<\mu_{\rm min}$ as a function of $\mu_{\rm min}$ for our simulated galaxies in the rest-frame UV (top) and B band (bottom) at $z=6$. We show the results for galaxies at several intrinsic magnitudes in $-18<\muv<-12$ and $-18<\rm M_B<-12$ (averaged over all simulated galaxies at a given magnitude in our sample). Our calculation indicates that at the limits of HFF (26.5\,mag\,arcsec$^{-2}$) and HUDF ($\sim1$\,mag deeper than HFF), more than 80--90\% of the rest-frame UV light from galaxies brighter than $\muv<-18$ should be detected, but this fraction is much smaller for fainter galaxies. In the rest-frame B band, a larger fraction of the light is in low surface brightness regions, as expected from the fact that B-band light is more spatially extended than the UV. Figure \ref{fig:flight} provides information on what depths the observations need to reach for certain targets, although in practice one also needs to account for the PSF of the observational facilities for quantitative comparison.

\subsection{The nature of UV-bright clumps}
\label{sec:clump}
Our simulations suggest that $z\geq5$ galaxies are mostly irregular, with rest-frame UV images dominated by a few bright clumps. These clumps mainly have two different origins: some of them are satellite galaxies falling on to their host (e.g. galaxy C in Figure \ref{fig:image}), while others are groups of young stars formed collectively from a parent cloud, i.e. massive giant molecular cloud-like complexes (galaxies D and F). The latter is similar to the clumps formed in gas-rich disks via disk instabilities in intermediate-redshift massive galaxies in simulations \citep[e.g.][]{hopkins.2012:clumpy.disk,genel.2012:giant.clump.in.sims,moody.2014:clump.simulation,oklopcic.2017:fire.giant.clump,mandelker.2017:giant.clumps} and observations \citep[e.g.][]{guo.2015:clumpy.galaxy.candels}. These high-redshift galaxies are gas-rich and highly turbulent, in part due to rapid accretion from the intergalactic medium. The high degree of turbulent support causes the gas to fragment into large clumps, which subsequently form stars. These early galaxies often do not have well-defined, rotationally supported disks.. 

The two formation channels mentioned above are essentially the same as the {\em ex-situ} and {\em in-situ} clumps defined in \citet{mandelker.2017:giant.clumps}. Many clumps formed `{\em in-situ}' are dynamically short-lived (as seen at intermediate-redshift galaxies in \citealt{oklopcic.2017:fire.giant.clump}). For example, the brightest clump in galaxy F (also see the top-right panel in Figure \ref{fig:szmu}) contains a mass of $2\times10^5\,\Msun$ in stars within 100\,pc (central surface density $\sim50\,\Msun\,\pc^{-2}$) that are formed simultaneously 6\,Myr ago; the clump is unbound with a virial parameter $\alpha_{\rm vir}\sim2E_{\rm k}/|E_{\rm p}|\sim10$, and it will be dispersed to $\sim500\,\pc$ in size within $\sim30\,\Myr$. However, these simulations also form long-lived bound stellar clumps that survive more than 400 Myr, after which the present simulations end. Some of these stellar clumps might survive and evolved into present-day globular clusters \citep{kim.2018:globular.candidate.fire}. Bound cluster are more likely to form once the initial gas surface density exceeds $\sim500\,\Msun\,\pc^{-2}$ \citep[also see][]{grudic.2017:high.sf.efficiency}. 

Finally, we caution that these UV-bright clumps are observationally `short-lived': they become much fainter after 30\,Myr as the light-to-mass ratio decreases by more than a factor of 10 following stellar evolution and the loss of massive stars. Even the dynamically long-lived clumps are difficult to identify at later times if they only contribute a small fraction of the total stellar mass. Consequently, the rest-frame UV morphology and size of a galaxy can vary greatly on $\sim30\,\Myr$ time-scale due to stellar evolution, even if the stellar mass morphology and size do not change dramatically.

\section{Conclusions}
\label{sec:conclusion}
In this paper, we use high-resolution FIRE-2 cosmological zoom-in simulations to predict galaxy morphologies and sizes during the epoch of reionization. We project the star particles onto a two-dimensional grid to make stellar surface density and UV and B-band surface brightness images, and measure the half-mass and half-light radii in UV and B band for our simulated galaxies at $z=5$--10. Our main findings are as follows: 

(i) The simulated galaxies show a variety of morphologies at similar magnitude and/or stellar mass, from compact galaxies to clumpy, multi-component galaxies to irregular galaxies. The rest-frame UV images are dominated by a few bright, small young stellar clumps that are often not always associated with a large stellar mass. The rest-frame B-band images are determined both by the bulk of stars and by the bright clumps (Section \ref{sec:overview} and Figure \ref{fig:image}). 

(ii) At any redshift, there is a correlation between galaxy size and stellar mass/luminosity with large scatter. At fixed stellar mass, the half-mass radius spans over a factor of 5, while at fixed magnitude, the half-light radius (both UV and B band) spans over a factor of 20 from less than 0.01" up to 0.2" (Figure \ref{fig:szms}).

(iii) Galaxy morphologies and sizes in our simulations depend on the band in which they are observed. Going from the intrinsic stellar mass distribution to rest-frame B band to rest-frame UV, galaxies appear smaller and more concentrated. (Figure \ref{fig:szsz}). The half-mass radii correlate with B-band half-light radii better than those in the UV, suggesting that B-band light is a better tracer of stellar mass than the UV light, but it can also be strongly biased by the UV bright clumps.

(iv) At $z\geq5$, the physical sizes of galaxies at fixed stellar mass and/or magnitude decrease with increasing redshift as $(1+z)^{-m}$ with $m\sim1$--2 (Figure \ref{fig:sz}). For galaxies below $\Ms\sim10^8\,\Msun$, the ratio of the half-mass radius to the halo virial radius is $\sim10\%$ and does not evolve at $z=5$--10. More massive galaxies have smaller stellar-to-halo size ratios, typically 1--5\% (Section \ref{sec:evolution}).

(v) The observed half-light radius of a galaxy strongly depends on the surface brightness limit of the observational campaign (Figure \ref{fig:szmu}). This effect may account for the extremely small galaxy sizes and size--luminosity relation measured in the Hubble Frontier Fields observations (Figure \ref{fig:hff}), as shallower observations can be dominated by single young stellar `clumps'. We provide the cumulative light distribution of surface brightness for typical $z=6$ galaxies (Figure \ref{fig:flight}).


In this paper, we make predictions to help understand current and plan future observations of faint galaxies at $z\geq5$. Our prediction that these galaxies have small, bright clumps on top of more extended, low surface brightness regions can be tested in the near future by high-resolution deep imaging with JWST on a typical sample of galaxies. In future work, we intend to make more realistic comparisons with specific observational campaigns to understand the sample completeness and their implications for the faint-end UV luminosity functions. We will also study the size evolution for a broad range of galaxies from $z=0$--10 (Schmitz et al., in preparation). Moreover, it is also worth quantifying the statistical and physical properties of the UV-bright clumps in $z\geq5$ galaxies.

\section*{Acknowledgments}
We acknowledge the anonymous referee for useful comments that help improve this manuscript. 
We thank Rychard Bouwens, Brian Siana, James Bullock, and Eros Vanzella for helpful discussion. 
The simulations used in this paper were run on XSEDE computational resources (allocations TG-AST120025, TG-AST130039, TG-AST140023, and TG-AST140064). 
The analysis was performed on the Caltech compute cluster ``Zwicky'' (NSF MRI award \#PHY-0960291).
Support for PFH was provided by an Alfred P. Sloan Research Fellowship, NASA ATP Grant NNX14AH35G, and NSF Collaborative Research Grant \#1411920 and CAREER grant \#1455342.
MBK acknowledges support from NSF grant AST-1517226 and from NASA grants NNX17AG29G and HST-AR-12836, HST-AR-13888, HST- AR-13896, and HST-AR-14282 from the Space Telescope Science Institute, which is operated by AURA, Inc., under NASA contract NAS5-26555.
CAFG was supported by NSF through grants AST-1412836, AST-1517491, AST-1715216 and CAREER award AST-1652522, by NASA through grant NNX15AB22G, and by STScI through grant HST-AR-14562.001.
EQ was supported by NASA ATP grant 12-APT12-0183, a Simons Investigator award from the Simons Foundation, and the David and Lucile Packard Foundation.
RF acknowledges financial support from the Swiss National Science Foundation (grant no 157591).
SGK was supported by NASA through Einstein Postdoctoral Fellowship grant number PF5-160136 awarded by the Chandra X-ray Center, which is operated by the Smithsonian Astrophysical Observatory for NASA under contract NAS8-03060.
DK was supported by NSF grant AST-1412153, funds from the University of California, San Diego, and a Cottrell Scholar Award from the Research Corporation for Science Advancement. 
AW was supported by NASA through grants HST-GO-14734 and HST-AR-15057 from STScI.

\bibliography{ms}

\appendix

\section{Particle smoothing and size measurements}
\label{sec:append}

\begin{figure}
\centering
\includegraphics[width=\linewidth]{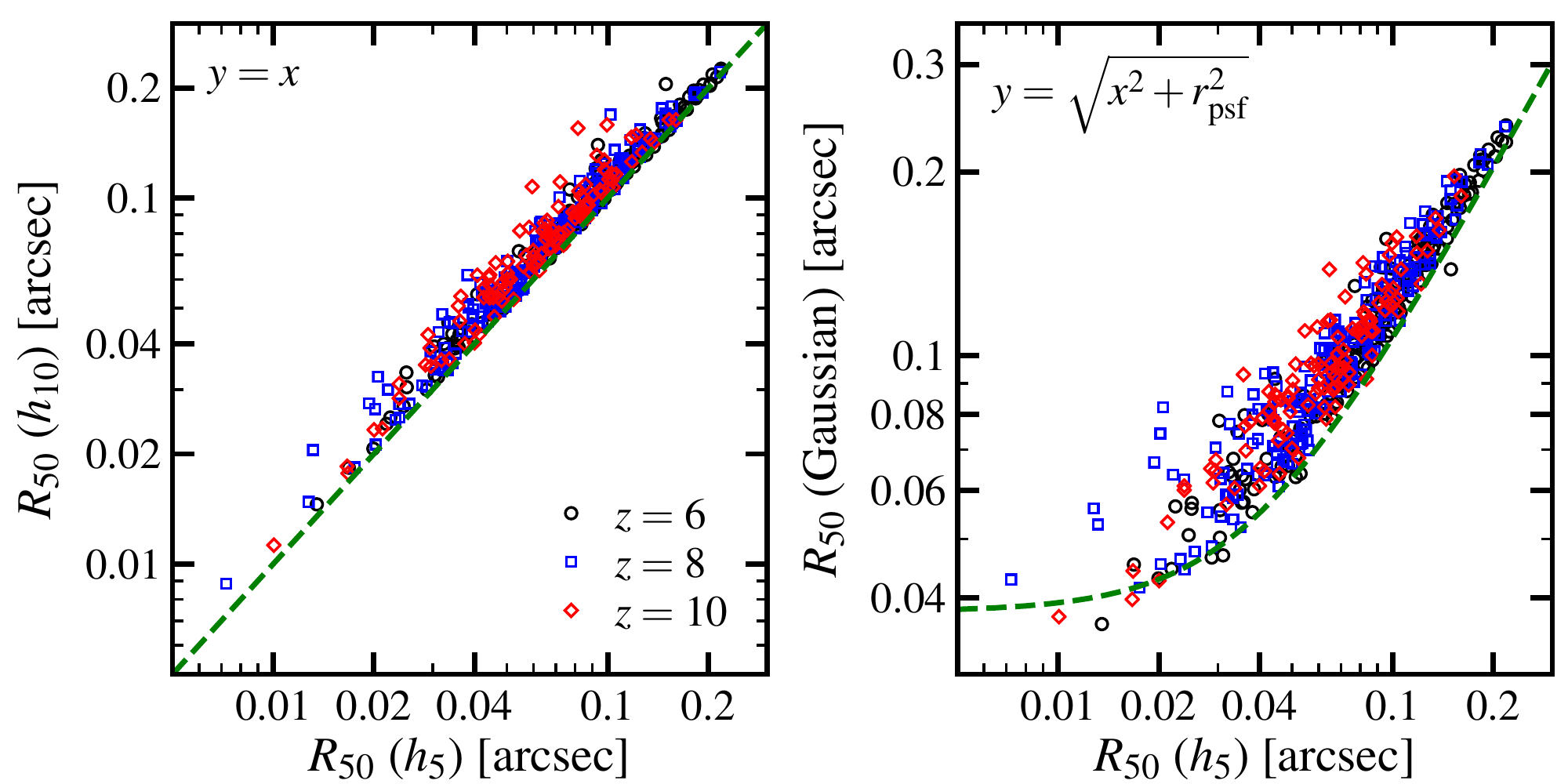}
\caption{Galaxy sizes measured using different smoothing approaches (see text for details). Our default size measurements are reasonably numerically robust to this choice.}
\label{fig:rcomp}
\end{figure}

In this work, we adopt a non-parametric approach to define galaxy half-mass (light) radii by measuring the area spanned by the brightest pixels that contribute 50\% of the total intensity within an 1"-diameter, $S_{50}$, and taking $R_{50}=\sqrt{S_{50}/\pi}$ (Section \ref{sec:definition}). We note that the results weakly depend on how we smooth the star particles on the projected images. By default, each star particle is smoothed over a cubic spline kernel with a smoothing length $h_5$ equal to its distance to the $5^{\rm th}$ nearest particle. Here we discuss two alternative smoothing approaches. First, we adopt a smoothing length $h_{10}$ (the distance to the $10^{\rm th}$ nearest particle) instead of $h_5$ (but still use the cubic spline kernel) and repeat the size measurement. In the left panel of Figure \ref{fig:rcomp}, we compare the new half-mass radii with our default results for our simulated galaxies. The green dashed line shows the $y=x$ relation. By using $h_{10}$, the half-mass radii only increase by less than 10\% for most of the galaxies (5\% difference on average), and only a small fraction (1\%) of our galaxies are affected by 50\% or more.

Alternatively, we further smooth our default images using a two-dimensional Gaussian function with a dispersion corresponding to the size of 10 pixels (0.032"). This equals to the pixel size of the NIRCam on JWST. This is to mimic the observed galaxy images after convolving with the PSF. A point source thus has a half-mass (light) radius of $r_{\rm psf}=0.038$". Note that the example images in Figure \ref{fig:image} are generated in this way for better visualization. We repeat the non-parametric size measurement on the Gaussian-smoothed images and compare the results in the right panel of Figure \ref{fig:rcomp}. The green dashed line shows the $y=\sqrt{x^2+r_{\rm psf}^2}$ relation for reference: we note that this relation is also used in observations to convert apparent sizes to intrinsic sizes \citep[e.g.][]{oesch.2010:hudf.morphology}. Nearly all of our simulated galaxies lie close to this curve (less than 20\% deviation) as expected. 

These experiments suggest that we obtain numerically stable galaxy half-mass radii by using a cubic spline kernel with smoothing length $h_5$. We find similar results for half-light radii in UV and B band: using $h_{10}$ instead, the B-band sizes are not affected by more than 10\% for the vast majority of our galaxies. The differences are slightly larger for UV sizes. 5\% of our galaxies have UV half-light radii increased by a factor of 1.5--2 when using $h_{10}$. This is because the UV light in these galaxies is dominated by few diffuse star particles that have large inter-particle distance, so the sizes we obtain can only be treated as upper limits. Nonetheless, most galaxies in our sample are only affected by less than 20\%. We conclude that our non-parametric size measurement is robust to our particle smoothing method.

\section{Resolution convergence}
\label{sec:append:res}
\begin{figure}
\centering
\includegraphics[width=0.9\linewidth]{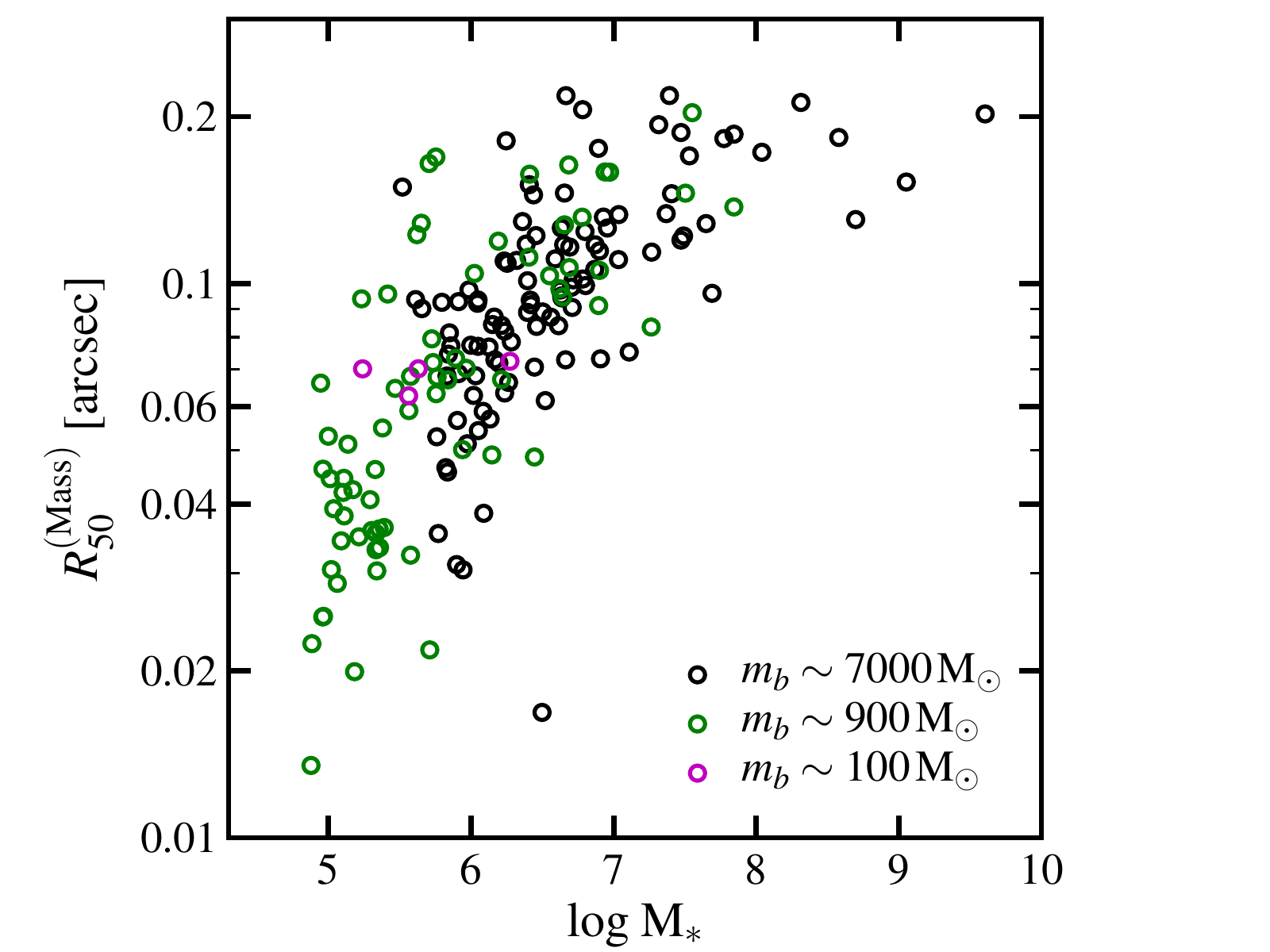}
\caption{The $z=6$ galaxy size--mass relation. Colors show simulations run with different mass resolution. Galaxy sizes converge reasonably well with resolution in our simulations.}
\label{fig:res}
\end{figure}

We note that our sample includes simulations using three different mass resolutions for baryonic particles ($\mb\sim100$, 900, and 7000\,$\Msun$). We showed in previous papers that galactic scale quantities, such stellar mass, star formation rates, etc., converge reasonably well at these mass resolutions \citep[e.g.][]{hopkins.2017:fire2.numerics,ma.2017:fire.hiz.smf}. Here in Figure \ref{fig:res}, we show the $z=6$ galaxy size--mass relation for our simulated sample, where the colors represent simulations run with different mass resolution. There is no significant difference between different resolution levels in the size--mass relation, so we conclude that galaxy sizes are robust with respect to resolution in our simulations.

\label{lastpage}

\end{document}